# User Defined Spreadsheet Functions in Excel

Dermot Balson
Perth, Australia

Jerzy Tyszkiewicz
Institute of Informatics, University of Warsaw
Banacha 2, 02-097 Warszawa, Poland
jty@mimuw.edu.pl

**ABSTRACT**

*Creating user defined functions (UDFs) is a powerful method to improve the quality of computer applications, in particular spreadsheets. However, the only direct way to use UDFs in spreadsheets is to switch from the functional and declarative style of spreadsheet formulas to the imperative VBA, which creates a high entry barrier even for proficient spreadsheet users. It has been proposed to extend Excel by UDFs declared by a spreadsheet: user defined spreadsheet functions (UDSFs).*

*In this paper we present a method to create a limited form of UDSFs in Excel without any use of VBA. Calls to those UDSFs utilize what-if data tables to execute the same part of a worksheet several times, thus turning it into a reusable function definition.*

## 1 INTRODUCTION

Defining procedures and functions is a key ingredient of structured programming. It allows for creating extensive libraries consisting of high quality components, which are fundamental in rapid development of software systems.

Excel as a software system is a bundle of a functional, declarative language of spreadsheet formulas and an imperative language of macros – Visual Basic for Applications (VBA). They are fundamentally different and require completely different styles of programming, so that proficiency in one of them does not help in learning the other. In fact, many experienced spreadsheet users avoid using VBA.

Taking those factors into account, it is not surprising that [Peyton Jones et al. 2003] proposed an extension to Excel by allowing user defined functions defined by means of spreadsheets. They wrote: "*From a programming language point of view, then, spreadsheets lack the most fundamental mechanism that we use to control complexity: the ability to define re-usable abstractions. In effect, they deny to end-user programmers the most powerful weapon in our armory.*"

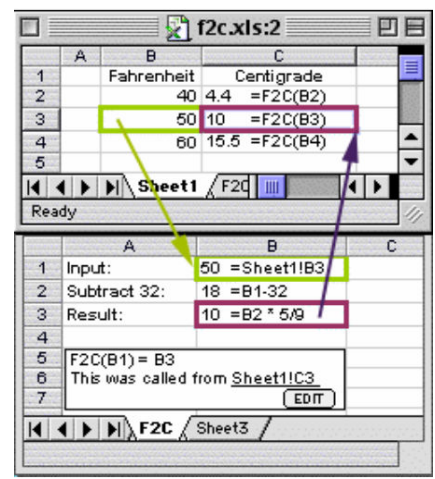

**From [Peyton Jones et al. 2003]**

Later, they identify the main target users of the proposed solution to be the moderate users, i.e."*[…] those who understand the spreadsheet paradigm fairly thoroughly. Not only have they already mastered the prerequisites, but they also tackle more ambitious and long-lived applications […]*"

Then the authors conclude: "*The implementation of a function must be defined by a spreadsheet, because that is the only computational paradigm understood by our target audience.*"

However, this extension has not been incorporated into Excel editions 2007 and 2010, despite favorable evaluation by the prospective users.

The main aim of the present paper is to offer good news: **It is possible to create and call user defined spreadsheet functions in Excel.** We call them UDSFs henceforth, to distinguish them from user defined functions written in VBA, commonly abbreviated as UDFs.

Since UDSFs are not natively supported in Excel, our solution expresses them by (ab)using another functionality of this spreadsheet system: what-if data tables, which were available already back in 2003, when [Peyton Jones et al. 2003] was written.

We hope that present research will

- offer an improvement to the present state of spreadsheet programming, which led [Panko 2008] to the bitter conclusion that "*In general, end user development in spreadsheeting seems to resemble programming practice in the 1950s and 1960s*";

- be a complement to [Grossman et al. 2009] who analyzed the technique of using lookup functions in place of nested IFs in spreadsheets, demonstrating how a complicated programming task can be better realized by a spreadsheet function intended for a completely different purpose;

- be a simpler and more transparent way to realize the idea of *repositories of spreadsheet components*, which were really intended to be equivalents of subroutine libraries, except that no way to call those subroutines was known and therefore the solution relied on automated transformations of the components to allow their insertion into the spreadsheet under development [Paine 2008];

- be a vehicle to bring the so-much-needed modularization into spreadsheets [Paine 2005];

- offer a method to realize really complex data processing tasks in a transparent and understandable way, e.g. SQL queries, which can be implemented in spreadsheets according to [Tyszkiewicz 2010].

However, we do not intend to propose a methodology of incorporating UDSFs into business practice already now. We present and test the technical properties of the implementation of this feature, leaving that other task for the future research.

The outline of the paper is as follows:

- In Section 2 we describe ISBN – the standard example in this paper.

- In Section 3 we describe the data table construction of UDSF calls.

- In Section 4 we discuss many technical issues of using UDSF calls, including parameter passing, construction of function libraries, constructing columns with many calls, speed of computation and side effects.

- In the final Section 5we present concluding remarks.
- Bibliography is Section 6.

## 2 EXAMPLE: ISBN NUMBERS

We describe an example of a possible application of UDSFs. It is related to ISBN, the widely known numbering systems for books, defined as international standard ISO 2108:1978.

The digits of ISBN-13 are divided into 5 blocks of variable length:

1. GS1 prefix 978;
2. the group (language) identifier;
3. the publisher code;
4. item number;
5. single control digit

Codes of the older ISBN-10 standard miss the GS1 prefix, and in the remaining part have the same structure: 10 digits in four blocks. Its control digit definition is different, in particular letter X can be used as the control digit. It is still in common use due to longevity of books. Therefore any validation procedure for ISBN should handle both ISBN-13 and ISBN-10.

Verifying that the control digit is valid in an ISBN-13 number is rather straightforward. For ISBN-10 it is a little more complicated, since this ISBN must be encoded as text (in order to accommodate X and possibly handle leading 0s). More comprehensive (and more complicated) tests are also possible, verifying the validity of the group identifier and publisher code by looking up tables of already assigned values.

We refer the reader to [Wikipedia 2012] for more information, including the definition of the control digit generation algorithms (access to the official ISO documents is paid and expensive). In general, the non-control digits of the ISBN are multiplied by certain weights and added, and this value determines the control digit.

ISBNs can be stored in many possible forms in a spreadsheet: as a texts of length 10 or 13, as four or five adjacent cells containing texts representing the blocks, or 10 to 13 cells containing separate digits. Checking the validity of an ISBN number stored in a spreadsheet will be the running example used in this paper.

## 3 USER DEFINED SPREADSHEET FUNCTIONS VIA DATA TABLES

### 3.1 Technical Introduction

The technical issue of defining a UDSF to check the validity of an ISBN (let's name it *ISBNcheck*) by a fragment of spreadsheet is quite simple, if we intend to use *ISBNcheck* only once. We can create a separate worksheet *ISBNcheck*for that purpose, identify the cells intended to be inputs and assign a name to the output (let it be named ISBNcheck). Then in a location where we want to call *ISBNcheck* we insert =ISBNcheck, link the input cells of *ISBNcheck* to the cells which hold the inputs we want to call *ISBNcheck* with and finished. A separate part of the spreadsheet is computing *ISBNcheck*.

The problems begin when we want to call *ISBNcheck* several times with different inputs. This means, that the worksheet *ISBNcheck* should be evaluated many times, each time with different values fed to its input cells. This section presents a method to achieve that effect using *what-if data tables*, a tool introduced into spreadsheets for completely different purpose.

Let us start with the formulas for validating the ISBN on the control digit level (due to their length, we will not reproduce them any more).

|   | A | B | C | D |
|---|---|---|---|---|
| 1 | ISBN | ISBN10 | ISBN13 | RESULT |
| 2 | 8320425395 | valid | #VALUE! | valid |

| B2 | =IF(12-MOD(SUMPRODUCT(VALUE(MID(A2,{1;2;3;4;5;6;7;8;9},1)),{10;9;8;7;6;5;4;3;2}),11)<br>=MATCH(RIGHT(A2),{"0";"1";"2";"3";"4";"5";"6";"7";"8";"9";"X"},0),"valid","invalid") |
|---|---|
| C2 | =IF(MOD(<br>10-MOD(SUMPRODUCT(VALUE(MID(A2,{1;2;3;4;5;6;7;8;9;10;11;12},1)),{1;3;1;3;1;3;1;3;1;3;1;3}),10),10)<br>=VALUE(RIGHT(A2)),"valid","invalid") |
| D2 | =IF(ISBLANK(A2),"",IF(LEN(A2)=10,B2,C2)) |

### 3.2 What-If Data Tables

Data tables are part of the what-if analysis tools present in Excel. Below we describe the construction of a one input data table, using ISBN validation as an example.

|   | A | B | C | D |
|---|---|---|---|---|
| 1 | ISBN | ISBN10 | ISBN13 | RESULT |
| 2 |  |  | #VALUE! | #VALUE! |
| 3 |  |  |  |  |
| 4 |  |  |  |  |
| 5 | 0201038013 | valid |  |  |
| 6 | 0201038021 | valid |  |  |
| 7 | 020103803X | valid |  |  |
| 8 | 9780201134476 | valid |  |  |
| 9 | 8320425395 | valid |  |  |

| B4 | =D2 |
|---|---|
| B5 | {=TABLE(,A2)} |
| B6 | {=TABLE(,A2)} |
| B7 | {=TABLE(,A2)} |
| B8 | {=TABLE(,A2)} |
| B9 | {=TABLE(,A2) |

Formulas in row 2 are the same as before.

In order to verify five ISBNs we select the range A4:B9 (outlined on the figure above) and convert it into a *one input data table* by going through a standard dialog. The key ingredient is that during the dialog we specify the *input cell* A2, which is fed with the value of the argument and initiates the computation which actually verifies the validity of an ISBN code. The formula in B4 specifies where the result to be displayed in the data table is located. In order to evaluate the data table, the following procedure is repeated for each row: the ISBN candidate from the left column is fed into A2 (the left red arrow), the evaluation of all dependent cells is carried out (indicated by the blue dependence arrows drawn by Excel), and the result from D2 is inserted (via B4) into the right column in the same row (two red arrows on the right). Then the next row is processed in the same manner, and so on. Finally, the original content of A2 is restored and dependent cells recomputed: A2isempty in our example, which causes two formulas in row 2to return error values (which are harmless in our situation), as well as D2 and B4to become blank.

**3.3 Data Table UDSF Calls**

Crucially, if there are many data tables in one worksheet which share the same input cell, and thus are being computed by the same fragment of spreadsheet, the system schedules the computations and executes them sequentially, one by one.

This mechanism makes data tables a suitable tool to create UDSF calls. Indeed, each 2 by 2 data table may be considered as a single call to UDSF*ISBNcheck*, like in the example below.

|    | A          | B       | C        | D       |
|----|------------|---------|----------|---------|
| 1  | ISBN       | ISBN10  | ISBN13   | RESULT  |
| 2  |            | #VALUE! | #VALUE!  |         |
| 3  |            |         |          |         |
| 4  |            |         |          |         |
| 5  | 8320425395 | valid   |          |         |
| 6  |            |         | 0201038021 | valid |
| 7  |            |         |          |         |
| 8  | 9780201038099 | valid |         |         |
| 9  |            |         | 020103803X | valid |
| 10 |            |         |          |         |
| 11 | 0201038013 | valid   |          |         |
| 12 |            |         |          |         |

| B4  | =D2          |
|-----|--------------|
| B5  | {=TABLE(,A2)} |
| D5  | =D2          |
| D6  | {=TABLE(,A2)} |
| B7  | =D2          |
| B8  | {=TABLE(,A2)} |
| D8  | =D2          |
| D9  | {=TABLE(,A2)} |
| B10 | =D2          |
| B11 | {=TABLE(,A2)} |

Formulas in row 2 are the same as before.

The computation is carried out outside of the call location, and the Excel mechanisms take care of executing the calls one by one. The function body is indicated by bold border, each separate call is in thin border, the actual call results have gray background.

This functionality is completely automated and the calls may be located in an arbitrary way. However, the array formulas with TABLE function cannot be inserted directly into the cell, the user must go through the what-if analysis menu. Worse still, in Excel they cannot even be copied or filled (only cut and pasted), and thus there is no simple way (except using VBA) to insert many such calls is a worksheet.

**4. DESIGNING AND CALLING USER DEFINED SPREADSHEET FUNCTIONS**

**4.1 Argument Passing**

The limitation to only 1 argument cell may be a problem for data table UDSF calls (there are two input data tables, as well, but this is still very little). Additionally, it is impossible to pass directly ranges as arguments.

Both of those obstacles can be circumvented by using *call by reference* method of passing arguments to UDSFs. This notion comes from the theory of programming languages and means the mechanism of calling a function when the argument for the functions is not a value (say: number), but the memory address where that number can be found. Excel provides a few methods to specify addresses of cells and retrieve the values from them. It can be a textual reference (to be used with INDIRECT), or a number (coordinate of the cell in a known range, to be used with OFFSET or INDEX). Each of those descriptions can be passed to the procedure and allow it to retrieve the number from that cell.

Let us describe this on our ISBN example. For that purpose, let us assume that ISBN-10 codes are not stored in single cells, but in four cells each, one block per cell.

In this situation, the argument to the call is the text "A5:D5" representing the range, where the arguments are located.

|   | A | B | C | D | E | F |
|---|---|---|---|---|---|---|
| 1 | INPUT ADDRESS | ISBN | ISBN10 | RESULT | | |
| 2 | | #REF! | #REF! | | | |
| 3 | | | | | | |
| 4 | | | | | | |
| 5 | 0 | 201 | 03803 | X | A5:D5 | valid |

| | |
|---|---|
| B2 | =INDEX(INDIRECT(A2),1)&INDEX(INDIRECT(A2),2)&INDEX(INDIRECT(A2),3)&INDEX(INDIRECT(A2),4) |
| D2 | =IF(ISBLANK(A2),"",C2)) |
| F4 | =E2 |
| E5 | ="A5:D5" |

Function *ISBNcheck* has been modified accordingly: the potential ISBN is reconstructed in cell B2 using INDIRECT, and then the formulas in row 2(omitted in the table) validates it. D2 produces the final result, since we do not intend to validate ISBN-13 numbers. Of course, this function can now be conveniently extended to do a more comprehensive validation, since it has access to the blocks of the ISBN, which are not easy to isolate otherwise, because their lengths are not fixed.

The same method can be used to pass ranges to UDSFs, if we wish them to accept arguments of this type.

### 4.2 Library Construction and Use

Most users dealing with ISBNs in spreadsheets will implement the most rudimentary check, if any. It would be a significant benefit for them if they could get a workbook implementing the optimal, fully comprehensive validation, because probably very few of them have time and necessary resources to create it themselves. Such a workbook could also contain validation formulas for other common identifiers, like ISSN, ISMN, ISRC, etc.

We describe now how such a workbook can be used as a *library of UDSFs* for rapid development of a new workbook.

So let us assume that such a workbook saved as lib.xlsx is available, with one sheet for each function provided by the library (let us assume it contains two functions: *ISBN10check* and *ISSNcheck*). The sheet for *ISBN10check* is named ISBN10check and has a clearly marked input cell(say A2) and a clearly marked output cell (say B9). It contains all formulas necessary for computing *ISBN10check*, and extensive instructions, including information how the input to *ISBN10check* should be created, what is the recommended way of passing arguments to it, etc.

Now let us assume the user creates a new workbook Book2 and wants to insert a call to *ISBN10check* in its worksheet Sheet1. The user opens lib.xlsx in addition to the already created new workbook. First, the library sheet with the function definition and the newly created sheet must be linked together. This is achieved as follows:

1) In Sheet1 the user reserves an input cell to *ISBN10check*. Let it be A1. (Data tables require that their input cells are located in the same sheet.)

2) The user enters =[Book2]Sheet1!A1 into the sheet ISBN10check!A2 of lib.xlsx.

At this moment *ISBN10check* and Sheet1 are linked and the function can be called in the sheet. Identical steps can be undertaken to link function *ISSNcheck* to Sheet1, if desired.

Let's now describe how a call to *ISBN10check* is constructed.

3) The user creates a 2 by 2 data table with column input cell A1. In the top-right cell of the newly created data table he/she enters =[lib]ISBN10check!B9 and the call is ready. Now it is only necessary to enter the argument to the call in the lower left call of the call. As many calls can be constructed as necessary, the same input cell can serve all of them simultaneously.

Calls to *ISSNcheck* can be created in the same worksheet, they do not interfere with the previous calls.

If *ISBN10check* (or *ISSNcheck*) from the same library should be called in another worksheet Sheet2, the user may either create a new copy of ISBN10check and repeat the procedure above, or modify the input link in ISBN10check located in ISBN10check!A1 to the following formula:

4) =IF(ISBLANK([Book2]Sheet2!A1),[Book2]Sheet1!A1,[Book2]Sheet2!A1)

At this moment *ISBNcheck* is already linked simultaneously to Sheet1 and Sheet2 and can be called in both of them. The calls need not be altered. This procedure can obviously be generalized for more worksheets.

| Cell | Book2.xlsx |
|---|---|
| F2 | =[lib]ISBN10check!B9 |
| F3 | {=TABLE(,A1)} |
| | lib.xlsx |
| A2 | =[Book2]Sheet1!A1 |
| B2 | =INDEX(INDIRECT(A2),1) |
| B3 | =INDEX(INDIRECT(A2),2) |
| B4 | =INDEX(INDIRECT(A2),3) |
| B5 | =INDEX(INDIRECT(A2),4) |
| B6 | =B2&B3&B4 |
| B7 | =12-MOD(SUMPRODUCT(VALUE(MID(B6,{1;2;3;4;5;6;7;8;9},1)),{10;9;8;7;6;5;4;3;2}),11) |

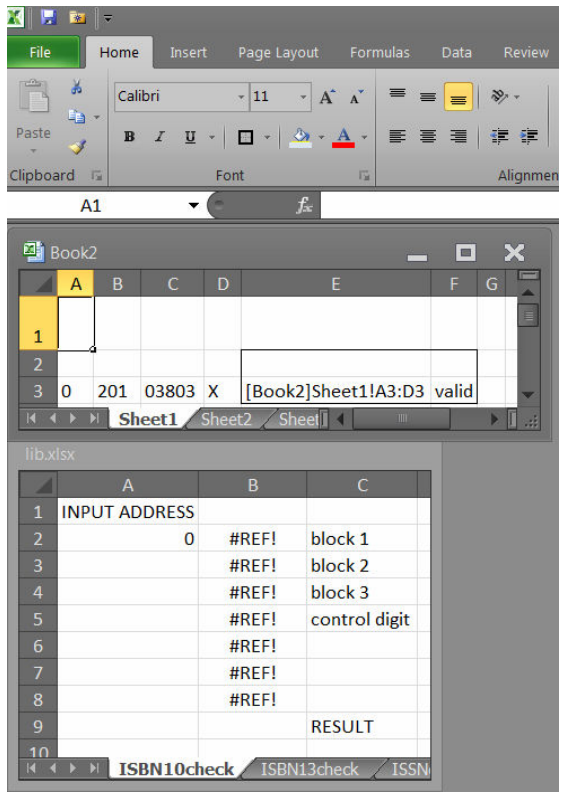

The overall picture of the situation: the library workbook and the new workbook open simultaneously in Excel are shown on the figure, the table shows all formulas. The argument [Book2]Sheet1!A3:D3 has been created by manually. The layout resembles the picture from [Peyton Jones et al. 2003], where UDSFs have been first proposed, indicating that we have constructed in Excel what the authors wanted to add to it.

The worksheets with function definitions can be left in the library workbook, as long as it remains open during computation. If the user wishes to close the library workbook, the worksheets with UDFS definitions should be moved to Book2 by drag-and-drop.

The question arises, why the user might want to use this method, which contradicts the usual habits one develops learning spreadsheet programming. The answer is two-fold. On the one hand, it seems unlikely to us, that the users will start creating UDSFs and modularizing spreadsheets by themselves – a good deal of training is necessary to learn that. One the other hand, if libraries of carefully developed and tested, high quality UDSFs become available, many programmers might decide to learn how to use them rather than write the functionalities they need from scratch, and likely get inferior results anyway.

**4.4 Columns and Rows of Calls to User Defined Spreadsheet Functions**

Data tables can be copied neither by copy-paste nor by filling in Excel. Therefore it is difficult to create a large number of them automatically. Fortunately, there is a simple workaround allowing one to create columns and rows of calls to a single UDSF. The solution is to use a larger data table in place of many smaller ones. A careful look at picture in Section 3.2 reveals that we indeed have a column of 5 calls to the same UDSF there. Rows of UDSF calls can be created using data tables with inputs in columns.

An additional benefit of this organization is that a single large data table is computed much faster than a large number of individual small data tables, which execute each of the calls separately.

UDSF calls are volatile, hence they are recomputed after every change in the worksheet, unless the user switches automatic data table recomputation off in the Excel options. In this case recomputation must be enforced manually. To make things still worse, a worksheet with data table UDSF calls is executed by Excel using only one thread, irrespectively of how many of them are available. This slows down the execution in many cases. Finally, the computation of many separate UDSF calls is slow because of the high overhead associated with processing of data tables.

The tests reported below were conducted in Excel 2010 running on a laptop with Intel Core i3-2330M

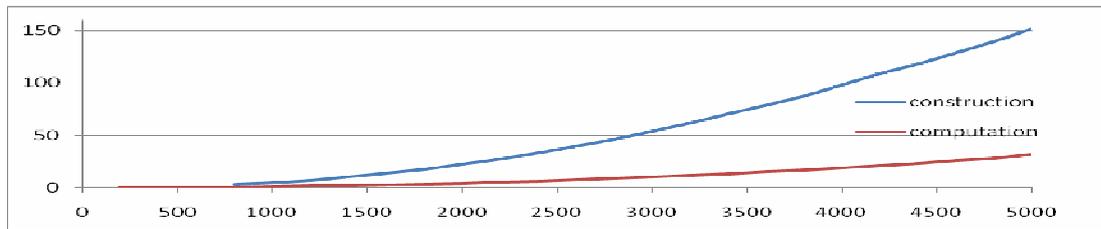

CPU @2.20 GHz (2 cores, 4 hardware threads), 3 GB RAM under Windows 7 Professional 32 bit with SP1. In all tests, the computation was performed by only one thread – apparently Excel is unable to parallelize computations of data tables.

The chart above shows how fast separate UDSF calls of the standard size 2 by 2 are created and evaluated. The functions used were extremely simple, so that the measured times are almost entirely due to the processing of data tables.

We recorded a simple macro adding to a single worksheet one by one 100 calls to a function located in another worksheet, and 100 calls to another similar function in a separate worksheet, and then recomputing the worksheet. The vertical axis is time of adding those 200 calls measured in seconds and the time to compute all calls in the worksheet. The horizontal axis shows the number of already existing calls. As one may notice, adding a single call to 5000 already existing ones takes about 1 second.

Then we made 2 additional copies of the worksheet with calls, which resulted in 15000 calls in total. Surprisingly enough, copying went very fast, almost instantly. The recomputation of the whole workbook took over 310 seconds, more than 5 minutes.

We also tested the speed of computation of UDSF calls located in a column and implemented by a single large data table. We created a workbook containing calls in columns. Each of 3 worksheets contained two data tables with 10000 calls to the same two functions as before each. Thus in total there were 60000 calls. Recomputation of this worksheet took about 0.5 second. This indicates the tremendous efficiency difference between computing many small data tables and a few large ones.

**4.5 Side Effects**

An important property of functional languages, in particular of the language of spreadsheet formulas, is that functions do not have *side effects*. This means that a call to function does not change anything in the system except that the value of the function is computed (and possibly triggers the recomputation of other formulas, which depend on it). UDSFs do not have side effects, if iterative calculations are switched off in the Excel options. If they are on, it is possible to create a side effect of a UDSF call. However, even if iterative calculations are on, side effects seem unlikely to be caused accidentally.

**4.6 Nesting and Recursion**

Unfortunately, it turns out that UDSFs cannot be nested. Experiments have shown that a call to a UDSF located inside the body of another UDSF is indeed not re-evaluated when the body of that function is evaluated during an external call. Therefore library functions must be constructed without using UDSFs.

For the same reason, the mechanism of UDSFs cannot implement recursion, either.

### 4.7 Spreadsheet Auditing

UDSF calls cannot be traced by built-in auditing tools of Excel, which do not reveal that their outputs depend on the inputs. No workaround seems to be possible.

One may inspect the body of the function which is called and audit how it is evaluated for each particular input. This can be achieved by manually inserting the inputs into the input cells of the function. This does not affect its operation.

### 4.8 Other Spreadsheet Systems

Data tables are not supported in OpenOffice, LibreOffice and Google docs, so this form of UDSF calls is impossible in those systems.

Data tables are supported in gnumeric, work very well and can be copied and filled, unlike in Excel. In this respect gnumeric seems to be a perfect tool to make use of this mechanism. The bad news is that files exported from gnumeric in Excel-readable formats are either corrupt (xlsx, ods) or not fully functional (xls –input cells of data tables are relocated to the same position where the data table itself is located, which effectively makes them useless).

## 5. CONCLUSIONS

### 5.1 Summary

- Data tables seem to be a very promising method to implement UDSFs in Excel.

- They offer the possibility of creating and using libraries of functions in Excel, a crucial tool in software engineering, which speeds up creation of programs and reduces the rate of programming errors in the whole application. We hope that emergence of such libraries might lure spreadsheet programmers to start using UDSFs.

- UDSF calls generally slow down computations by preventing Excel from using multiple threads and by introducing overheads caused by the data table processing. Therefore they will probably be more important in rapid development of small and mid-size spreadsheets, where the time of computing is not a crucial factor. They might also be used in rapid prototyping of larger applications, with the option to be replaced by standard spreadsheet calculations later on, to get a more effective solution – perhaps using the methods described in [Paine 2008].

- A big disadvantage is that a single UDSF call occupies an inconvenient 2 by 2 range and cannot be copied or filled in Excel. However, a simple macro can be used to insert data tables. Creating large ranges of such calls can be achieved by using larger data tables instead of many separate calls.

### 5.2 Excel Development Suggestions

- It seems very desirable that Excel is extended by data tables in other, more convenient formats, together with a mechanism of copying and filling them. It should not be very complicated, since the computation engine is already capable of handling them and it seems that only interface issues must be resolved. However, there are many other factors involved here: e.g., this functionality would be incompatible with the earlier versions of Excel.

- Improvements to the auditing functions are needed, to let them track how data table entries depend on the inputs. Optimization of data table computations would be very welcome.

- Since *call by reference* is important in data tables (see Section 4.2), it would be beneficial to introduce another form of ADDRESS function (say XADR), which would convert its argument range to its textual address (including worksheet and workbook name), e.g. `=XADR(A1:A5)` used in `Sheet1` of `Book2` should return `[Book2]Sheet1!A1:A5` as text. It would be the optimal way to create arguments of UDSF calls. At present this can be semi-automatically (one has to specify manually if the reference is absolute/relative and the worksheet and workbook name) achieved by the formula

    `="[Book2]"&ADDRESS(ROW(A1:A5),COLUMN(A1:A5),4,,"Sheet1")&":"&`

    `ADDRESS(ROW(A1:A5)+ROWS(A1:A5)-1,COLUMN(A1:A5)+COLUMNS(A1:A5)-1,4)`

    but it would be beneficial to have something easier to use and harder to make a mistake in. In Section 4.2 we used a manually created address.

## ACKNOWLEDGMENTS

This paper has benefited substantially from the suggestions of two anonymous reviewers, whom we would like to express our sincere thanks. The research of J.T. has been funded by Polish National Science Centre (Narodowe Centrum Nauki)